\documentclass[hyper]{JHEP3}

\title{Hypercharge and baryon minus lepton number in $E_6$}
\author{Junpei Harada\\ 
        Department of Physics, Hiroshima University, 
        Higashi-Hiroshima 739-8526, Japan \\
        E-mail: \email{harada@theo.phys.sci.hiroshima-u.ac.jp}}

\abstract{
We study assignments of the hypercharge and baryon minus lepton number 
for particles in the $E_6$ grand unification model.
It is shown that there are three assignments of hypercharge
and three assignments of baryon minus lepton number 
which are consistent with the
Standard Model. Their explicit expressions and detailed properties are given.
In particular, we show that the $U(1)_{B-L}$ symmetry in $E_6$ cannot be
orthogonal to the $SU(3)_R$ symmetry. 
Based on these investigations, 
we propose an alternative $SU(5)$ grand unification model.}


\begin{document}

\section{\label{sec:level1}Introduction}
The supersymmetric (SUSY) grand unified theory (GUT) is one of the most 
attractive scenarios beyond the Standard Model (SM).
Among all the possible SUSY GUT models,
the minimal SUSY $SU(5)$ model has been most extensively studied in the past.
However, recent experimental progress suggests that it is no longer
the primary candidate and some modifications are obviously required for
several reasons. The first reason is that the lower limit on the proton
lifetime measured by the Super-Kamiokande~\cite{SK99} 
already excludes the minimal 
SUSY $SU(5)$ model~\cite{MP01}~\footnote{There is still controversy in
this statement.
This issue is discussed in detail in refs.~\cite{BPS02}\cite{EW03}.}.
The second reason is that small neutrino masses suggested by recent
neutrino oscillation experiments~\cite{nu-exp} can be naturally
explained by the seesaw mechanism~\cite{seesaw}, which requires
right-handed neutrinos in addition to the minimal SM particles.

Although an extension to some 
non-minimal SUSY $SU(5)$ GUT models is one possible
solution, it is worth investigating GUT models with alternative
gauge groups. 
Extending the gauge group beyond $SU(5)$ is interesting 
since the theory includes the right-handed neutrinos and we have 
several ways of embedding the SM gauge group into the extended gauge group.
For example, in the $SO(10)$ GUT there are two different assignments of
hypercharge, which are related with each other by the $SU(2)_R$
symmetry. This fact leads to two different GUT models 
in the context of the $SU(5)$ group;
the Georgi-Glashow model~\cite{GG74} or the flipped
model~\cite{Barr82}~\cite{flipped}.
Recent studies show that the SUSY flipped $SU(5)$ model overcomes the 
difficulties in the minimal SUSY $SU(5)$ GUT~\cite{Reflipped}.

The $E_6$ is an unique exceptional Lie group that has complex representations
and gives an observed chiral structure at low-energies.
The $E_6$ GUT~\cite{E6} has been known to give, for example, 
the small Cabbibo-Kobayashi-Maskawa (CKM) mixing angles and bi-large
neutrino mixings~\cite{BK99}\cite{E6model}. 
Since $E_6$ is larger than
$SO(10)$, there may be other different assignments of hypercharge and 
baryon minus lepton number $(B-L)$ 
in the $E_6$ grand unification. 
These possibilities lead to new alternative GUT models.

In this paper 
we thoroughly investigate assignments of the
hypercharge and $B-L$ in the $E_6$ unification.
We show that in the $E_6$ GUT there are 
{\it three} assignments of hypercharge ~\cite{Hyper} and 
{\it three} assignments of $B-L$ that reproduce the SM. 
Their properties are very important for model building.
We point out that these three assignments are related 
with each other by the $SU(2)$ subgroup of $SU(3)_R$, which is the
subgroup of $E_6$.
We also show that each assignment is
orthogonal to the $SU(2)$ subgroup of $SU(3)_R$.
In particular, we emphasize that the $U(1)_{B-L}$ symmetry in $E_6$ can not
be orthogonal to the $SU(3)_R$ symmetry. 
This fact strongly restricts the $E_6$ grand unified theories
with the gauged $U(1)_{B-L}$ symmetry.
In those of $9(3 \times 3)$ pairs of charge assignments, 
we show that $6$ pairs are consistent with the SM.
These observation indicate
that $E_6$ has much potential for constructing
alternative GUT models compared to the $SO(10)$ GUT.

We also propose a $E_6$-inspired alternative $SU(5)$ GUT model, 
``E-twisting flipped model''.
The gauge group of this model is $SU(5)\times U(1)_V \times
U(1)_{V^\prime}$ which is the subgroup of $E_6$.
Assignments of hypercharge and $B-L$ 
are different from those of the Georgi-Glashow model
or usual flipped model. In this model, since the hypercharge is not a
subgroup of $SU(5)$, the quasi-unification of strong, weak
and hypercharge gauge couplings is predicted.
While Dirac masses of quarks or leptons 
are naturally obtained through the renormalizable operators 
in the superpotential,
the heavy Majorana masses of right-handed neutrinos are obtained 
via dimension five operators. 
Therefore, the conventional seesaw mechanism can work.

This paper is organized as follows. In section $2$ we review
$SO(10)$ GUT since it will be helpful when discussing $E_6$.
Especially, we emphasize that as far as $SO(10) \supset
U(1)_{B-L}$ is imposed, only the ${\bf 16}$ representation, not ${\bf 10}$, 
can be a candidate for the SM matters in both the Georgi-Glashow model
and usual flipped model.
In section $3$ we investigate the 
hypercharge and $B-L$ assignments in the $E_6$ unification. 
We show that there are three assignments of hypercharge or 
$B-L$.
Their explicit expressions and detailed properties are given.
In section $4$, as an example, 
we propose a $E_6$-inspired alternative $SU(5)$ GUT model. 
Section $5$ is devoted to the conclusion.

\section{\label{sec:level1}Particle assignment in $SO(10)$}
A crucial difference between the $SO(10)$ GUT and the $SU(5)$ GUT 
resides in the fact that the way of embedding the SM gauge group 
into the unified gauge group is unique or not. 
In the $SU(5)$ GUT it is unique, while in the $SO(10)$ GUT 
there are two different assignments of hypercharge. 
Since different assignments lead to different predictions,
this problem is important for low energy physics.
In this section we study this issue in detail.

In the minimal $SO(10)$ GUT, all matters in the SM 
and a right-handed neutrino
belong to ${\bf 16}$ for each family.
Each spinor representation ${\bf 16}$ is decomposed
under a maximal subgroup 
$SO(10) \supset SU(5) \times U(1)_V$ as
\begin{eqnarray}
 {\bf 16} = {\bf 5^*}_{3} + {\bf 10}_{-1} + {\bf 1}_{-5},
\end{eqnarray}
and furthermore under $SU(5) \supset SU(3)_C \times SU(2)_L \times U(1)_Z$ as
\begin{eqnarray}
 {\bf 5^*}& = & ({\bf 3^*},{\bf 1})_{1/3} + ({\bf 1},{\bf 2})_{-1/2} , 
                \nonumber \\
 {\bf 10} & = & ({\bf 3},{\bf 2})_{1/6} + ({\bf 3^*},{\bf 1})_{-2/3} 
               +({\bf 1},{\bf 1})_1 , \nonumber \\
 {\bf 1}  & = & ({\bf 1},{\bf 1})_0.
\end{eqnarray}
Notice here that $U(1)_Z$, which is the subgroup of $SU(5)$, is not identical
with the $U(1)_Y$ hypercharge at this stage. Clearly, each
${\bf16}$ includes two pairs of $({\bf 3^*},{\bf 1})$ or $({\bf 1},{\bf 1})$. 
This fact can be understood by considering the decomposition of {\bf 16}
under another maximal subgroup 
$SO(10) \supset SU(4)_{PS} \times SU(2)_L \times SU(2)_R$,
\begin{eqnarray}
 {\bf 16} = ({\bf 4},{\bf 2},{\bf 1}) + ({\bf 4^*},{\bf 1},{\bf 2}),
\end{eqnarray}
and under $SU(4)_{PS} \times SU(2)_L \times SU(2)_R \supset 
SU(3)_C \times SU(2)_L \times SU(2)_R \times U(1)_X$ as
\begin{eqnarray}
 ({\bf 4},{\bf 2},{\bf 1}) & = & ({\bf 3},{\bf 2},{\bf 1})_{1/3}
                                +({\bf 1},{\bf 2},{\bf 1})_{-1} ,
                                 \nonumber \\
 ({\bf 4^*},{\bf 1},{\bf 2}) & = & ({\bf 3^*},{\bf 1},{\bf 2})_{-1/3}
                                +({\bf 1},{\bf 1},{\bf 2})_{1}.
\end{eqnarray}
This decomposition shows that two pairs of $({\bf 3^*},{\bf 1})$ or
$({\bf 1},{\bf 1})$ are $SU(2)_R$ doublets. 
One can also find that the quantum numbers of 
$U(1)_Z$ and $U(1)_V$ are represented by the one of $U(1)_X$ and the third
component $I_{3R}$ of $SU(2)_R$ as follows,
\begin{eqnarray}
 V & = & -4I_{3R} -3X , \nonumber \\
 Z & = & -I_{3R} + \frac{1}{2}X.
\end{eqnarray}
Note that the $X$-charge can also be identified with $B-L$,
\begin{eqnarray}
 B - L = X = -\frac{1}{5}(V-4Z).
\end{eqnarray}
Therefore, the
$U(1)_{B-L}$ symmetry in $SO(10)$ is orthogonal to the $SU(2)_R$ symmetry.

From above equations, one can find that 
$SU(5)$ multiplets ${\bf 5^*}$,${\bf 10}$ and 
${\bf 1}$ are formed as 
\begin{eqnarray}
 {\bf 5^*}_3 & = & ({\bf 3^*},{\bf 1}, I_{3R} = -1/2)_{-1/3}
                  +({\bf 1},{\bf 2},{\bf 1})_{-1} , \nonumber \\
 {\bf 10}_{-1} & = & ({\bf 3},{\bf 2},{\bf 1})_{1/3}
                    +({\bf 3^*},{\bf 1},I_{3R}=1/2)_{-1/3}
                    +({\bf 1},{\bf 1},I_{3R}=-1/2)_{1} ,\nonumber \\
 {\bf 1}_{-5} & = & ({\bf 1},{\bf 1},I_{3R} = 1/2)_1.
\end{eqnarray}
These arguments indicate 
that the way of embedding $SU(5)$ as $SO(10) \supset
SU(5) \supset SU(3)_C \times SU(2)_L$ is not unique~\cite{BK99}.  
As far as the $U(1)_Y$ hypercharge is not considered,
there is freedom of $SU(2)_R$ rotation.
It is worth mentioning here that there still remains 
$SU(2)_R$ freedom of embedding $SU(5)$ as
$SO(10) \supset SU(5) \times U(1)_V \supset SU(3)_C \times SU(2)_L
\times U(1)_{B-L}$. This is based on the following relation; 
$U(1)_{B-L} \perp SU(2)_R$.

Let us consider the hypercharge assignments in the $SO(10)$ GUT. 
The hypercharge $U(1)_Y$ is not identical with $U(1)_X$ and 
must be orthogonal to $SU(3)_C$
and $SU(2)_L$. This fact shows that the $U(1)_Y$ hypercharge is not
orthogonal to $SU(2)_R$. In other words,
the assignment of the hypercharge eliminates the $SU(2)_R$ freedom.

In order to give explicit expressions for the hypercharge,
consider the following decomposition 
$SO(10) \supset SU(5)\times U(1)_V \supset 
 SU(3)_C \times SU(2)_L \times U(1)_Z \times U(1)_V$. 
Therefore, 
the hypercharge $U(1)_Y$ must be a linear combination of $U(1)_Z$ and
 $U(1)_V$; $U(1)_Y \subset U(1)_Z \times U(1)_V$.
There are two assignments of hypercharge that reproduce the 
SM~\cite{Barr82},
\begin{enumerate}
 \item $\frac{Y}{2} = Z$ ; Georgi-Glashow model~\cite{GG74}
 \item $\frac{Y}{2} = -\frac{1}{5}(Z+V)$ ; flipped
       model~\cite{Barr82}~\cite{flipped}.
\end{enumerate}
These hypercharge assignments can also be expressed in terms of the third
component of $SU(2)_R$ and the quantum number of $U(1)_X$ as
\begin{eqnarray}
 \frac{Y}{2} & = & Z
               = -I_{3R} + \frac{1}{2}X,
\end{eqnarray}
for the Georgi-Glashow model and 
\begin{eqnarray}
 \frac{Y}{2} & = & -\frac{1}{5}(Z+V)
               =  I_{3R} + \frac{1}{2}X,
\end{eqnarray}
for the flipped model~\footnote{These expressions are useful for comparing two
different hypercharge assignments. The formula between hypercharge and
$B-L$ can be always reduced to a conventional one $Y/2=I_{3R}+(B-L)/2$ by
redefining the third component of $SU(2)_R$.}.
These two assignments are different with each other
in only the sign of the third component of $SU(2)_R$.
This means that the $SU(5)$ group of flipped model is obtained
from that of Georgi-Glashow model by the $\pi$ rotation in 
$SU(2)_R$~\cite{BK99}. Namely the 
particle assignment of the flipped $SU(5)$ model is 
given from that of the Georgi-Glashow $SU(5)$ model
by the interchange of $SU(2)_R$ doublets (flipping),
\begin{eqnarray}
 u^c \leftrightarrow d^c, \quad e^c \leftrightarrow \nu^c.
\end{eqnarray}
Although the way of embedding $SU(5)$ as $SO(10) \supset SU(5) \times U(1)_V
\supset SU(3)_C \times SU(2)_L \times U(1)_Y$ is not unique, there 
remains no $SU(2)_R$ freedom anymore. 
Therefore, there are only two possibilities;
$SU(5)$ must be $SU(5)_{GG}$ or $SU(5)_{flipped}$.

Finally, we investigate ${\bf 10}$ representations of $SO(10)$.
As mentioned above,
in the minimal $SO(10)$ GUT all matters in the SM and a right-handed neutrino
belong to
${\bf 16}$ for each family. 
However, since GUT models in which
${\bf 10}$ includes the SM matters are possible, 
it is worth investigating ${\bf 10}$ of $SO(10)$.
Each ${\bf 10}$ representation is decomposed under a maximal subgroup 
$SO(10) \supset SU(5) \times U(1)_V$ as
\begin{eqnarray}
 {\bf 10} = {\bf 5}_2 + {\bf 5^*}_{-2},
\end{eqnarray}
and under $SU(5) \supset SU(3)_C \times SU(2)_L \times U(1)_Z$ as
\begin{eqnarray}
 {\bf 5}   & = & ({\bf 3},{\bf 1})_{-1/3} + ({\bf 1},{\bf 2})_{1/2}, 
                 \nonumber \\
 {\bf 5^*} & = & ({\bf 3^*},{\bf 1})_{1/3} + ({\bf 1},{\bf 2})_{-1/2}.
\end{eqnarray}
As before, consider the decomposition of ${\bf 10}$ under another maximal
subgroup $SO(10) \supset SU(4)_{PS}\times SU(2)_L \times SU(2)_R$ as
\begin{eqnarray}
 {\bf 10} = ({\bf 6},{\bf 1},{\bf 1}) + ({\bf 1},{\bf 2},{\bf 2}),
\end{eqnarray}
and under $SU(4)_{PS}\times SU(2)_L \times SU(2)_R \supset 
SU(3)_C \times SU(2)_L \times SU(2)_R \times U(1)_X$ as
\begin{eqnarray}
 ({\bf 6},{\bf 1},{\bf 1}) & = & ({\bf 3},{\bf 1},{\bf 1})_{-2/3}
                                +({\bf 3^*},{\bf 1},{\bf 1})_{2/3} ,
                                 \nonumber \\
 ({\bf 1},{\bf 2},{\bf 2}) & = & ({\bf 1},{\bf 2},{\bf 2})_0. 
\end{eqnarray}
From these equations,
one can see that $SU(5)$ multiplets ${\bf 5}$ and ${\bf 5^*}$ are
formed as
\begin{eqnarray}
 {\bf 5}_2 & = & ({\bf 3},{\bf 1},{\bf 1})_{-2/3}
                +({\bf 1},{\bf 2},I_{3R}= -1/2)_0 , \nonumber \\
 {\bf 5^*}_{-2} & = & ({\bf 3^*},{\bf 1},{\bf 1})_{2/3}
                     +({\bf 1},{\bf 2},I_{3R}= 1/2)_0.
\end{eqnarray}
This indicates that if ${\bf 5^*}$ of ${\bf 10}$ is regarded as the 
SM matters, $U(1)_X$ cannot be identified with $U(1)_{B-L}$.
We emphasize that as far as $SO(10) \supset
U(1)_{B-L}$ is imposed, ${\bf 10}$
of $SO(10)$ must not be a candidate of the SM matters
for both the Georgi-Glashow model and the flipped model. 
It is also noted
here that even if $SO(10) \supset U(1)_{B-L}$ is not imposed,
${\bf 10}$ can not be a candidate of the SM matters in
the flipped model~\footnote{Here we mean ``flipped model'' 
as the model that is composed 
of $SU(5)$ multiplets 
${\bf 5}^* = (u^c,e,\nu), {\bf 10}=(d^c,u,d,\nu^c)$ and ${\bf 1}=e^c$.}.

\section{\label{sec:level1}Particle assignment in $E_6$}
The $E_6$ GUT is one of the most attractive grand unified models
and it is important for model building to investigate
all the possible assignments of the hypercharge and $B-L$ in the $E_6$
unification. Although the $E_6$ GUT is more complicated than
the $SO(10)$ GUT since the rank of $E_6$ is larger, 
the argument is essentially the same as for the $SO(10)$ GUT.

In the $E_6$ GUT, all matters in the SM (with a right-handed neutrino) 
and exotic matters
belong to a fundamental representation {\bf 27} for each family. 
Each ${\bf 27}$ representation is
decomposed under a maximal subgroup 
$E_6 \supset SO(10) \times U(1)_{V^\prime}$ as follows
\begin{eqnarray}
 {\bf 27} = {\bf 16}_1 + {\bf 10}_{-2} + {\bf 1}_4.
\end{eqnarray}
These $SO(10)$ multiplets ${\bf 16}$, ${\bf 10}$ and ${\bf 1}$ are
decomposed under $SO(10) \supset SU(5) \times U(1)_V$ as
\begin{eqnarray}
 {\bf 16} & = & {\bf 10}_{-1} + {\bf 5^*}_3 + {\bf 1}_{-5}, \nonumber \\
 {\bf 10} & = & {\bf 5}_2 + {\bf 5^*}_{-2} , \nonumber \\
 {\bf 1}  & = & {\bf 1}_0.
\end{eqnarray}
The decomposition of $SU(5)$ multiplets under 
$SU(5) \supset SU(3)_C \times SU(2)_L \times U(1)_Z$ is given in the 
previous section. One sees that each ${\bf 27}$ includes
three pairs of 
$({\bf3^*},{\bf 1})$, $({\bf 1},{\bf 2})$ or $({\bf 1},{\bf 1})$
, where the numbers in parentheses are the dimensions of $SU(3)_C$ and
$SU(2)_L$ respectively.
This can be understood by considering the decomposition of ${\bf 27}$
under a maximal subgroup $E_6
\supset SU(3)_C \times SU(3)_L \times SU(3)_R$ as
\begin{eqnarray}
 {\bf 27} = ({\bf 3},{\bf 3},{\bf 1}) + ({\bf 3^*},{\bf 1},{\bf 3^*})
           +({\bf 1},{\bf 3^*},{\bf 3}),
\end{eqnarray}
and under $SU(3)_C \times SU(3)_L \times SU(3)_R \supset
SU(3)_C \times SU(2)_L \times SU(3)_R \times U(1)_{Y_L}$ as
\begin{eqnarray}
 ({\bf 3},{\bf 3},{\bf 1}) & = & ({\bf 3},{\bf 2},{\bf 1})_{1/2}
                                +({\bf 3},{\bf 1},{\bf 1})_{-1}, 
                                \nonumber \\
 ({\bf 3^*},{\bf 1},{\bf 3^*}) & = & ({\bf 3^*},{\bf 1},{\bf 3^*})_0,
                                \nonumber \\
 ({\bf 1},{\bf 3^*},{\bf 3}) & = & ({\bf 1},{\bf 2},{\bf 3})_{-1/2}
                                + ({\bf 1},{\bf 1},{\bf 3})_1.
\end{eqnarray}
These equations show that three pairs of $({\bf 3^*},{\bf 1})$,
$({\bf 1},{\bf 2})$ or $({\bf 1},{\bf 1})$ are $SU(3)_R$ triplets.

For later discussions, consider further decompositions under
$SU(3)_C \times SU(3)_L \times SU(3)_R \supset 
 SU(3)_C \times SU(2)_L \times SU(2)_{(R)}
 \times U(1)_{Y_L} \times U(1)_{Y_{(R)}}$(the meaning of parentheses for
 $''R''$ will be clarified soon) as
\begin{eqnarray}
 ({\bf 3},{\bf 3},{\bf 1}) & = & ({\bf 3},{\bf 2},{\bf 1})_{1/2,0}
                                +({\bf 3},{\bf 1},{\bf 1})_{-1,0}, 
                                 \nonumber \\
 ({\bf 3^*},{\bf 1},{\bf 3^*}) & = & 
 ({\bf 3^*},{\bf 1},{\bf 2})_{0,-1/2} + ({\bf 3^*},{\bf 1},{\bf 1})_{0,1},
                                 \nonumber \\
 ({\bf 1},{\bf 3^*},{\bf 3}) & = &
       ({\bf 1},{\bf 2},{\bf 2})_{-1/2,1/2} 
     + ({\bf 1},{\bf 2},{\bf 1})_{-1/2,-1} 
     + ({\bf 1},{\bf 1},{\bf 2})_{1,1/2} 
     + ({\bf 1},{\bf 1},{\bf 1})_{1,-1},
\end{eqnarray}
where the numbers in parentheses are the dimensions of $SU(3)_C$,
$SU(2)_L$ and $SU(2)_{(R)}$ respectively, and subscripts represent the
quantum numbers of $U(1)_{Y_L}$ and $U(1)_{Y_{(R)}}$ respectively. 
Notice here that there are three $SU(2)$ subgroups of $SU(3)_R$.
We define three $SU(2)$ subgroups of $SU(3)_R$ as
follows~\footnote{Note that the definition of $SU(2)$ subgroups of
$SU(3)_R$ is not unique. 
We point out that there is another useful definition and it
will be discussed later.};
$({\bf 16},{\bf 10},{\bf 3^*},{\bf 1})$ and 
$({\bf 16},{\bf 5^*},{\bf 3^*},{\bf 1})$ are $SU(2)_R$ doublets,
$({\bf 16},{\bf 10},{\bf 3^*},{\bf 1})$ and 
$({\bf 10},{\bf 5^*},{\bf 3^*},{\bf 1})$ are $SU(2)_R^\prime$ 
                                         doublets~\cite{Ma87},
$({\bf 16},{\bf 5^*},{\bf 3^*},{\bf 1})$ and 
$({\bf 10},{\bf 5^*},{\bf 3^*},{\bf 1})$ are $SU(2)_E$ doublets~\cite{BK99},
where the numbers in parentheses are the dimensions of
$SO(10),SU(5),SU(3)_C$ and $SU(2)_L$ respectively.  
The meaning of parentheses for $''R''$ is clear now, namely, 
$SU(2)_{(R)}$ means $SU(2)_R$, $SU(2)_R^\prime$ or $SU(2)_E$.

The quantum numbers of 
$U(1)_{V^\prime}$, $U(1)_V$ and $U(1)_Z$ can be
represented by those of $U(1)_{Y_L}$, $U(1)_{Y_{(R)}}$ 
and the third component $I_{3(R)}$ of $SU(2)_{(R)}$ as
\begin{eqnarray}
 V^\prime & = &  2 Y_L           -2 Y_R , \nonumber \\
 V        & = & -2 Y_L -4 I_{3R} -2 Y_R , \nonumber \\
 Z        & = & \frac{1}{3}Y_L - I_{3R} + \frac{1}{3} Y_R,
\end{eqnarray}
or 
\begin{eqnarray}
 V^\prime & = & 2 Y_L + 3 I_{3R}^\prime + Y_R^\prime , \nonumber\\
 V        & = & -2 Y_L + I_{3R}^\prime + 3 Y_R^\prime , \nonumber\\
 Z        & = & \frac{1}{3}Y_L - I_{3R}^\prime + \frac{1}{3}Y_R^\prime,
\end{eqnarray}
or 
\begin{eqnarray}
 V^\prime & = & 2 Y_L + 3 I_{3E} + Y_E , \nonumber\\
 V        & = & -2 Y_L + 5 I_{3E} - Y_E , \nonumber\\
 Z        & = & \frac{1}{3} Y_L - \frac{2}{3} Y_E.
\end{eqnarray}

From these expressions, the
$SU(5)$ multiplets, which belong to a single ${\bf 27}$
of $E_6$, are formed under 
$SU(3)_C \times SU(2)_L \times SU(2)_R \times U(1)_{Y_L} \times
U(1)_{Y_R}$ as
\begin{eqnarray}
 ({\bf 16},{\bf 5^*}) & = & 
           ({\bf 3^*},{\bf 1},{\bf 2}(I_{3R}=-1/2))_{0,-1/2}
          +({\bf 1},{\bf 2},{\bf 1})_{-1/2,-1} ,        \nonumber \\
 ({\bf 16},{\bf 10}) & = &
           ({\bf 3},{\bf 2},{\bf 1})_{1/2,0} 
          +({\bf 3^*},{\bf 1},{\bf 2}(I_{3R}=1/2))_{0,-1/2}  
          +({\bf 1},{\bf 1},{\bf 2}(I_{3R}=-1/2))_{1,1/2} ,      \nonumber \\
 ({\bf 16},{\bf 1}) & = & 
           ({\bf 1},{\bf 1},{\bf 2}(I_{3R}=1/2))_{1,1/2} ,       \nonumber \\
 ({\bf 10},{\bf 5}) & = & 
           ({\bf 3},{\bf 1},{\bf 1})_{-1,0}            
          +({\bf 1},{\bf 2},{\bf 2}(I_{3R}=-1/2))_{-1/2,1/2} ,   \nonumber \\
 ({\bf 10},{\bf 5^*}) & = & 
           ({\bf 3^*},{\bf 1},{\bf 1})_{0,1}       
          +({\bf 1},{\bf 2},{\bf 2}(I_{3R}=1/2))_{-1/2,1/2} ,    \nonumber \\
 ({\bf 1},{\bf 1}) & = &
           ({\bf 1},{\bf 1},{\bf 1})_{1,-1}.      \label{eq:E6_1}
\end{eqnarray}
and then under $SU(3)_C \times SU(2)_L \times SU(2)_R^\prime \times
U(1)_{Y_L} \times U(1)_{Y_R}^\prime$ as
\begin{eqnarray}
 ({\bf 16},{\bf 5^*}) & = & 
           ({\bf 3^*},{\bf 1},{\bf 1})_{0,1}            
          +({\bf 1},{\bf 2},{\bf 2}(I_{3R}^\prime=1/2))_{-1/2,1/2} ,
            \nonumber \\
 ({\bf 16},{\bf 10}) & = &
           ({\bf 3},{\bf 2},{\bf 1})_{1/2,0} 
          +({\bf 3^*},{\bf 1},{\bf 2}(I_{3R}^\prime=1/2))_{0,-1/2}  
          +({\bf 1},{\bf 1},{\bf 2}(I_{3R}^\prime=-1/2))_{1,1/2} 
          ,    \nonumber \\
 ({\bf 16},{\bf 1}) & = & 
           ({\bf 1},{\bf 1},{\bf 1})_{1,-1} ,                \nonumber \\
 ({\bf 10},{\bf 5}) & = & 
           ({\bf 3},{\bf 1},{\bf 1})_{-1,0}                
          +({\bf 1},{\bf 2},{\bf 2}(I_{3R}^\prime=-1/2))_{-1/2,1/2} 
          , \nonumber \\
 ({\bf 10},{\bf 5^*}) & = & 
           ({\bf 3^*},{\bf 1},{\bf 2}(I_{3R}^\prime=-1/2))_{0,-1/2} 
          +({\bf 1},{\bf 2},{\bf 1})_{-1/2,-1} ,             \nonumber \\
 ({\bf 1},{\bf 1}) & = &
           ({\bf 1},{\bf 1},{\bf 2}(I_{3R}^\prime=1/2))_{1,1/2},  
           \label{eq:E6_2}
\end{eqnarray}
and furthermore under $SU(3)_C \times SU(2)_L \times SU(2)_E \times
U(1)_{Y_L} \times U(1)_{Y_E}$ as
\begin{eqnarray}
 ({\bf 16},{\bf 5^*}) & = & 
           ({\bf 3^*},{\bf 1},{\bf 2}(I_{3E}=1/2))_{0,-1/2}      
          +({\bf 1},{\bf 2},{\bf 2}(I_{3E}=1/2))_{-1/2,1/2} ,   
               \nonumber \\
 ({\bf 16},{\bf 10}) & = &
           ({\bf 3},{\bf 2},{\bf 1})_{1/2,0} 
          +({\bf 3^*},{\bf 1},{\bf 1})_{0,1}            
          +({\bf 1},{\bf 1},{\bf 1})_{1,-1} ,               \nonumber \\
 ({\bf 16},{\bf 1}) & = & 
           ({\bf 1},{\bf 1},{\bf 2}(I_{3E}=-1/2))_{1,1/2} ,  
          \nonumber \\
 ({\bf 10},{\bf 5}) & = & 
           ({\bf 3},{\bf 1},{\bf 1})_{-1,0}            
          +({\bf 1},{\bf 2},{\bf 1})_{-1/2,-1} ,            \nonumber \\
 ({\bf 10},{\bf 5^*}) & = & 
           ({\bf 3^*},{\bf 1},{\bf 2}(I_{3E}=-1/2))_{0,-1/2}    
          +({\bf 1},{\bf 2},{\bf 2}(I_{3E}=-1/2))_{-1/2,1/2} ,  
           \nonumber \\
 ({\bf 1},{\bf 1}) & = &
           ({\bf 1},{\bf 1},{\bf 2}(I_{3E}=1/2))_{1,1/2}.  
           \label{eq:E6_3}
\end{eqnarray}
As far as $SU(3)_C \times SU(2)_L$ is concerned, there is $SU(3)_R$
freedom. The relations among three $SU(2)$
subgroups of $SU(3)_R$ are given above.

Now we consider the hypercharge assignments in the $E_6$ unification. 
The hypercharge $U(1)_Y$ must be a linear combination of 
$U(1)_Z$, $U(1)_V$ and $U(1)_{V^\prime}$; $U(1)_Y \subset U(1)_Z
\times U(1)_V \times U(1)_{V^\prime}$. 
There are {\it three} assignments of hypercharge
that are consistent with the SM~\cite{Hyper},
\begin{enumerate}
 \item $\frac{Y}{2} = Z$,
 \item $\frac{Y}{2} = -\frac{1}{5}(Z+V)$ ,
 \item $\frac{Y}{2} = -\frac{1}{20}(4Z-V-5V^\prime)$.
\end{enumerate}
The $U(1)_Y$ hypercharge for the first assignment 
or the second one is a subgroup of
$SO(10)$, while the last $U(1)_Y$ is a subgroup of $E_6$.
Their properties are very important for model building.
Therefore, we express three assignments of hypercharge 
in terms of the quantum numbers of $U(1)_{Y_L}$, $U(1)_{Y_{(R)}}$
and the third component of $SU(2)_{(R)}$ as
\begin{eqnarray}
 \frac{Y}{2} & = & Z , \nonumber \\
             & = & \frac{1}{3}Y_L - I_{3R} + \frac{1}{3}Y_R , \nonumber \\
             & = & \frac{1}{3}Y_L - I_{3R}^\prime + \frac{1}{3}Y_R^\prime 
                                                            , \nonumber \\
             & = & \frac{1}{3}Y_L - \frac{2}{3}Y_E,       \label{eq:Y1}
\end{eqnarray}
or 
\begin{eqnarray}
 \frac{Y}{2} & = & -\frac{1}{5}(Z+V) , \nonumber \\
           & = & \frac{1}{3}Y_L + I_{3R} + \frac{1}{3}Y_R , \nonumber \\
           & = & \frac{1}{3}Y_L - \frac{2}{3}Y_R^\prime ,   \nonumber \\
           & = & \frac{1}{3}Y_L - I_{3E} + \frac{1}{3}Y_E, \label{eq:Y2}
\end{eqnarray}
or
\begin{eqnarray}
 \frac{Y}{2} & = & -\frac{1}{20}(4Z-V-5V^\prime) ,\nonumber \\
           & = & \frac{1}{3}Y_L - \frac{2}{3}Y_R , \nonumber \\
           & = & \frac{1}{3}Y_L + I_{3R}^\prime + \frac{1}{3}Y_R^\prime , 
                                                  \nonumber \\
           & = & \frac{1}{3}Y_L + I_{3E} + \frac{1}{3}Y_E.  \label{eq:Y3}
\end{eqnarray}
These expressions show the properties of each assignment 
and the relations among them.
The first hypercharge assignment $U(1)_Y$ is {\it not}
orthogonal to $SU(2)_R$ or $SU(2)_R^\prime$,  but {\it is} $SU(2)_E$ ; 
$U(1)_Y \not\perp SU(2)_R$ or $SU(2)_R^\prime$, but $U(1)_Y \perp SU(2)_E$.
The second assignment is ; 
$U(1)_Y \not\perp SU(2)_R$ or $SU(2)_E$,
but $U(1)_Y \perp SU(2)_R^\prime$. The third assignment is ;
$U(1)_Y \not\perp
SU(2)_R^\prime$ or $SU(2)_E$, but $U(1)_Y \perp SU(2)_R$. 
Thus, we conclude that in the $E_6$ GUT
any assignments of hypercharge are orthogonal to
the $SU(2)$ subgroup of $SU(3)_R$.
One can also find the relations among three hypercharge
assignments. Since differences of three hypercharge assignments are
only in the sign of the third component of $SU(2)_{(R)}$, 
three assignments are related with each other by the $\pi$ rotation of
$SU(2)_{(R)}$. Notice here that although the hypercharge
assignment eliminates $SU(3)_R$ freedom, $SU(2)_{(R)}$ freedom still remain
at this stage. This is different from the $SO(10)$ GUT case.

Now we come to the quantum number $B-L$ in the $E_6$ unification.
In this paper we concentrate on the case in which the $U(1)_{B-L}$ symmetry
is included in $E_6$ as a subgroup.
First, consider the following subgroup 
$E_6 \supset SU(3)_C \times SU(3)_L \times SU(3)_R
\supset SU(3)_C \times SU(2)_L \times U(1)_{Y_L} \times SU(3)_R$.
From Eq.~($3.4$),
$U(1)_{B-L}$ must not be identical with
$U(1)_{Y_L}$. On the other hand, 
$U(1)_{B-L}$ must be orthogonal to $SU(3)_C$
and $SU(2)_L$. From these facts, we conclude 
that the $U(1)_{B-L}$ symmetry in $E_6$ must not be
orthogonal to the $SU(3)_R$ symmetry; 
\begin{eqnarray}
 U(1)_{B-L} \not\perp SU(3)_R.
\end{eqnarray}
This relation is valid as far as $E_6 \supset U(1)_{B-L}$ is imposed.
Therefore,
the remaining $SU(2)_{(R)}$ freedom disappears.
This strongly restricts $E_6$ grand unified models
with the gauged $U(1)_{B-L}$ symmetry.

If $E_6 \supset U(1)_{B-L}$ is imposed,
$U(1)_{B-L}$ must be a linear combination of $U(1)_Z$, $U(1)_V$
and $U(1)_{V^\prime}$; $U(1)_{B-L} \subset U(1)_Z \times U(1)_V \times
U(1)_{V^\prime}$.
There are {\it three} assignments of $B-L$ that reproduce the SM;
\begin{enumerate}
 \item $B-L = -\frac{1}{5}(V-4Z)$,
 \item $B-L = \frac{1}{20}(16Z + V + 5V^\prime)$,
 \item $B-L = -\frac{1}{20}(8Z + 3V -5V^\prime)$.
\end{enumerate}
The properties of these $B-L $ assignments are understood by 
expressing it in terms of the charges of $U(1)_{Y_L}$, $U(1)_{Y_{(R)}}$
and the third component of $SU(2)_{(R)}$ as
\begin{eqnarray}
 B-L & = & -\frac{1}{5}(V-4Z),                            \nonumber \\
     & = & \frac{2}{3}Y_L + \frac{2}{3}Y_R,               \nonumber \\
     & = & \frac{2}{3}Y_L - I_{3R}^\prime - \frac{1}{3}Y_R^\prime, \nonumber\\
     & = & \frac{2}{3}Y_L - I_{3E} - \frac{1}{3}Y_E,      \label{eq:B-L1}
\end{eqnarray}
or 
\begin{eqnarray}
 B-L & = & \frac{1}{20}(16Z + V + 5V^\prime),            \nonumber\\
     & = & \frac{2}{3}Y_L - I_{3R} - \frac{1}{3}Y_R,     \nonumber \\
     & = & \frac{2}{3}Y_L + \frac{2}{3}Y_R^\prime,       \nonumber \\
     & = & \frac{2}{3}Y_L + I_{3E} - \frac{1}{3}Y_E,     \label{eq:B-L2}
\end{eqnarray}
or
\begin{eqnarray}
 B-L & = & -\frac{1}{20}(8Z + 3V - 5V^\prime),           \nonumber \\
     & = & \frac{2}{3}Y_L + I_{3R} - \frac{1}{3}Y_R,     \nonumber \\
     & = & \frac{2}{3}Y_L + I_{3R}^\prime - \frac{1}{3}Y_R^\prime, \nonumber\\
     & = & \frac{2}{3}Y_L + \frac{2}{3}Y_E.              \label{eq:B-L3}
\end{eqnarray}
One can easily see that three $B-L$ assignments are
orthogonal to the $SU(2)$ subgroup of $SU(3)_R$ 
and relations among three
assignments are similar to the case of the hypercharge, namely, 
they are related with each other by the $\pi$ rotation of $SU(2)_{(R)}$.

Since there are three assignments of hypercharge and three assignments of
$B-L$, $9(3\times 3)$
pairs of charge assignment exist. 
In those of $9$ pairs, $6$ pairs are consistent with the SM.
Since the case in which the hypercharge and $B-L$ are
orthogonal to the {\it same} $SU(2)$ subgroup 
of $SU(3)_R$ is not consistent with
the SM, $3$ pairs of $U(1)_Y$ {\it and} $U(1)_{B-L} \perp SU(2)_{(R)}$ 
are removed.

We summarize the 6 pairs of charge assignment which are consistent with the
SM;
\begin{enumerate}
 \item $U(1)_Y \perp SU(2)_E$ and $U(1)_{B-L} \perp SU(2)_R$,
 \item $U(1)_Y \perp SU(2)_E$ and $U(1)_{B-L} \perp SU(2)_R^\prime$,
 \item $U(1)_Y \perp SU(2)_R^\prime$ and $U(1)_{B-L} \perp SU(2)_R$,
 \item $U(1)_Y \perp SU(2)_R$ and $U(1)_{B-L} \perp SU(2)_R^\prime$,
 \item $U(1)_Y \perp SU(2)_R$ and $U(1)_{B-L} \perp SU(2)_E$,
 \item $U(1)_Y \perp SU(2)_R^\prime$ and $U(1)_{B-L} \perp SU(2)_E$.
\end{enumerate}
We express particle assignments for each case in terms of $SO(10)$.
The quantum numbers of each field are given in the table~1.

The first assignment is the most familiar one;
\begin{eqnarray}
 {\bf 16} &=& (d^c+e+\nu) + (u^c + u + d + e^c) + \nu^c, \nonumber \\
 {\bf 10} &=& (D + E^c + N^c) + (D^c + E + N) ,          \nonumber \\
 {\bf 1}  &=& S.
\end{eqnarray}
The second assignment~\cite{Ma96} is given from the first one by the $\pi$
rotation in $SU(2)_E$;
\begin{eqnarray}
 {\bf 16} &=& (D^c + E + N) + (u^c + u + d + e^c) + S,   \nonumber \\
 {\bf 10} &=& (D + E^c + N^c) + (d^c+e+\nu) ,            \nonumber \\
 {\bf 1}  &=& \nu^c.
\end{eqnarray}
The third assignment is given from the first one by the $\pi$ rotation in
$SU(2)_R$, namely the flipped model; 
\begin{eqnarray}
 {\bf 16} &=& (u^c+e+\nu) + (d^c + u + d + \nu^c) + e^c, \nonumber \\
 {\bf 10} &=& (D + E + N) + (D^c + E^c + N^c) ,          \nonumber \\
 {\bf 1}  &=& S.
\end{eqnarray}
The fourth one is given from the third one by the $\pi$ rotation in
$SU(2)_E$;
\begin{eqnarray}
 {\bf 16} &=& (D^c + E^c + N^c) + (d^c + u + d + \nu^c) + S, \nonumber \\
 {\bf 10} &=& (D + E + N) + (u^c+e+\nu) , \nonumber \\
 {\bf 1}  &=& e^c.
\end{eqnarray}
The fifth one is given from the first one by the $\pi$ rotation in
$SU(2)_R^\prime$;
\begin{eqnarray}
 {\bf 16} &=& (d^c + E^c + N^c) + (D^c + u + d + S) + \nu^c, \nonumber \\
 {\bf 10} &=& (D + e + \nu) + (u^c+E+N) , \nonumber \\
 {\bf 1}  &=& e^c.
\end{eqnarray}
The sixth one is given from the fifth one by the $\pi$ rotation in
 $SU(2)_E$;
\begin{eqnarray}
 {\bf 16} &=& (u^c+E+N) + (D^c + u + d + S) + e^c, \nonumber \\
 {\bf 10} &=& (D + e + \nu) + (d^c + E^c + N^c) , \nonumber \\
 {\bf 1}  &=& \nu^c.
\end{eqnarray}

These possibilities certainly indicate that $E_6$ has much potential
for constructing new alternative GUT models. 

Finally, we comment on the electromagnetic symmetry $U(1)_{em}$.
In the SM, the electromagnetic symmetry $U(1)_{em}$ is, of course, 
a linear combination of $SU(2)_L$ and $U(1)_Y$; 
$U(1)_{em} \subset SU(2)_L \times U(1)_Y$. 
From Eqs.~(\ref{eq:Y1}) - (\ref{eq:B-L3}), one can find that
$Y/2 =  - I_{3R} + (B-L)/2 = - I_{3R}^\prime + (B-L)/2$ 
for the first assignment of the hypercharge,
$Y/2 =    I_{3R} + (B-L)/2 = - I_{3E}^\prime + (B-L)/2$ 
for the second one, and
$Y/2 =    I_{3R}^\prime + (B-L)/2 =  I_{3E}^\prime + (B-L)/2$ 
for the third one. 
These relations and Eqs.~(\ref{eq:B-L1}) - (\ref{eq:B-L3})
indicate $U(1)_Y \subset SU(2)_{(R)} \times U(1)_{B-L}$ and 
$U(1)_{B-L} \subset U(1)_{Y_L} \times U(1)_{Y_{(R)}}$.
This observation may indicate that the origin of the electromagnetic symmetry
$U(1)_{em}$ is $SU(3)_L \times SU(3)_R$; 
$U(1)_{em} \subset SU(2)_L \times U(1)_Y \subset SU(2)_L \times
SU(2)_{(R)} \times U(1)_{B-L} \subset SU(2)_L \times SU(2)_{(R)} \times
U(1)_{Y_L} \times U(1)_{Y_{(R)}} \subset SU(3)_L \times
SU(3)_R$~\footnote{This relation becomes more simple form by redefining
the $SU(2)$ subgroups of $SU(3)_R$; the $SU(2)$ subgroup which
is orthogonal to $U(1)_Y$ is $SU(2)_E$, the $SU(2)$ subgroup which is
orthogonal to $U(1)_{B-L}$ is $SU(2)_R$, 
and the remaining $SU(2)$ subgroup is $SU(2)_R^\prime$. 
Using this definition
the electromagnetic symmetry $U(1)_{em}$ can be written as
$U(1)_{em} \subset SU(2)_L \times U(1)_Y \subset SU(2)_L \times SU(2)_R
\times U(1)_{B-L} \subset SU(2)_L \times SU(2)_R \times U(1)_{Y_L} \times
U(1)_{Y_R} \subset SU(3)_L \times SU(3)_R$.}.

\begin{table}
\caption{\label{tab:QN}The quantum numbers of left-handed fields that belong to
 ${\bf 27}$ representation of $E_6$.}
\begin{tabular}{|c|ccccc|}
\hline\hline
 fields  & $SU(3)_C$   & $I_{3L}$ & $Y$     & $B-L$   & $Q_{em}$ \\
\hline
 $u$     & ${\bf 3}$   & $1/2$    & $1/3$   & $1/3$   & $2/3$    \\
 $d$     & ${\bf 3}$   & $-1/2$   & $1/3$   & $1/3$   & $-1/3$   \\
 $u^c$   & ${\bf 3^*}$ & $0$      & $-4/3$  & $-1/3$  & $-2/3$   \\
 $d^c$   & ${\bf 3^*}$ & $0$      & $2/3$   & $-1/3$  & $1/3$    \\
 $\nu$   & ${\bf 1}$   & $1/2$    & $-1$    & $-1$    & $0$      \\
 $e$     & ${\bf 1}$   & $-1/2$   & $-1$    & $-1$    & $-1$     \\
 $\nu^c$ & ${\bf 1}$   & $0$      & $0$     & $1$     & $0$      \\
 $e^c$   & ${\bf 1}$   & $0$      & $2$     & $1$     & $1$      \\
 $D$     & ${\bf 3}$   & $0$      & $-2/3$  & $-2/3$  & $-1/3$   \\
 $E^c$   & ${\bf 1}$   & $1/2$    & $1$     & $0$     & $1$      \\
 $N^c$   & ${\bf 1}$   & $-1/2$   & $1$     & $0$     & $0$      \\
 $D^c$   & ${\bf 3^*}$ & $0$      & $2/3$   & $2/3$   & $1/3$    \\
 $N$     & ${\bf 1}$   & $1/2$    & $-1$    & $0$     & $0$      \\
 $E$     & ${\bf 1}$   & $-1/2$   & $-1$    & $0$     & $-1$     \\
 $S$     & ${\bf 1}$   & $0$      & $0$     & $0$     & $0$      \\
\hline\hline
\end{tabular}
\end{table}

\section{\label{sec:level1}Model}
In this section, as a simple example, 
we propose a $E_6$-inspired $SU(5)$ GUT model, 
``E-twisting flipped model''. 
This model is based on the gauge group $SU(5)\times U(1)_V
\times U(1)_{V^\prime}$, which is a subgroup of $E_6$.
We also assume ${\cal N}=1$ supersymmetry.
In this model, the hypercharge $U(1)_Y$ is shared among $SU(5)$, $U(1)_V$ and
$U(1)_{V^\prime}$, and is explicitly given by
\begin{eqnarray}
 \frac{Y}{2} = -\frac{1}{20}(4Z -V -5V^\prime),
\end{eqnarray}
where notation is the same as in the previous section.
This hypercharge $U(1)_Y$
is orthogonal to the $SU(2)_R$ symmetry.
On the other hand, $B-L$ is given by 
\begin{eqnarray}
 B-L = \frac{1}{20}(16Z + V + 5V^\prime),
\end{eqnarray}
which is orthogonal to the $SU(2)_R^\prime$ symmetry.

All matters in the SM and a right-handed neutrino in a family belong to 
\begin{eqnarray}
F_{\bf 5^*} & = & {\bf 5^*}_{-2,-2} = (u^c,e,\nu) , \nonumber \\
T_{\bf 10}  & = & {\bf 10}_{-1,1}  = (d^c,u,d,\nu^c), \nonumber \\
X_{\bf 1}  & = & {\bf 1}_{0,4}   = e^c,
\end{eqnarray}
where subscripts are the quantum numbers of $U(1)_V$ and
$U(1)_{V^\prime}$ respectively.
At first glance, this particle assignments seems to be the same 
as those of usual flipped model.
However, it is quite different. The $U(1)_Y$ hypercharge and 
$B-L$ of our model are not
subgroups of $SO(10)$ such as usual flipped model, but of $E_6$.
Since this model is given from the usual flipped model by the $\pi$ rotation in
$SU(2)_E$ (E-twisting~\cite{BK99}), we call this model as ``E-twisting
flipped model''. 
It is worth mentioning here that while $SU(5)$ multiplets 
${\bf 5^*}$, ${\bf 10}$ and ${\bf 1}$ of
usual flipped model can form a single ${\bf 16}$ representation of $SO(10)$,
$SU(5)$ multiplets in our model cannot. They must be 
$({\bf 10},{\bf 5^*})$, $({\bf 16},{\bf 10})$ and $({\bf 1},{\bf 1})$,
where the numbers in the parentheses are the dimensions of $SO(10)$ and
$SU(5)$.
Therefore, we introduce the following exotic matters to cancel anomaly;
$\tilde{T}_{\bf 5^*} \equiv {\bf 5^*}_{3,1}$,
$\tilde{F}_{\bf 5}   \equiv {\bf 5}_{2,-2}$ and 
$\tilde{T}_{\bf 1}   \equiv {\bf 1}_{-5,1}$.
Although our model seems somehow complicated 
because of the existence of exotic matters, 
its structure is quite simple. 
Indeed, the SM matters and the exotic matters
can form a single ${\bf 27}$ representation of $E_6$.
The exotic matters will decouple at low-energies.
We assign the chiral superfields of the SM matters and exotic
matters the odd $Z_2$ parity to avoid unwanted operators. 
This $Z_2$ parity can be regarded as an extended
$R$-parity. Indeed, this extended $R$-parity is not identical with the 
conventional one $R=(-1)^{2S+3(B-L)}$, where $B,L$ and $S$ are the
baryon, lepton and spin quantum numbers.

GUT symmetry breaking $SU(5)\times U(1)_V \times U(1)_{V^\prime}
\rightarrow SU(3)_C \times SU(2)_L \times U(1)_Y$ is induced by Higgs
multiplets
$\Sigma_{\bf 10}\equiv{\bf 10}_{-1,1}$ and $\overline{\Sigma}_{\bf
10^*}\equiv{\bf 10^*}_{1,-1}$, 
where the electrically neutral components develop large vacuum 
expectation values (VEV),
$\langle \nu^c_\Sigma \rangle = \langle \nu^c_{\overline{\Sigma}}
\rangle \equiv M$.
On the other hand, the Higgs multiplets 
which give Dirac masses to the SM matters are 
$\overline{\chi}_{\bf 5^*} \equiv {\bf 5^*}_{3,1}$ and 
$H_{\bf 5} \equiv {\bf 5}_{2,-2}$.
Notice here that $\overline{\chi}_{\bf 5^*}$ and $H_{\bf 5}$ should be 
$({\bf 16},{\bf 5^*})$ and $({\bf 10},{\bf 5})$ respectively.
This is different from the case of Georgi-Glashow model or usual flipped
model.
We introduce other Higgs multiplets to cancel anomaly;
$\phi_{\bf 1} \equiv {\bf 1}_{-5,1}$,
$\overline{\phi}_{\bf 1} \equiv {\bf 1}_{5,-1}$,
$\chi_{\bf 5} \equiv {\bf 5}_{-3,-1}$,
$\overline{H}_{\bf 5^*} \equiv {\bf 5^*}_{-2,-2}$,
$H^\prime_{\bf 5} \equiv {\bf 5}_{2,2}$ and 
$\overline{H^\prime}_{\bf 5^*} \equiv {\bf 5^*}_{-2,2}$.
We assign $Z_2$ parity even to the Higgs chiral superfields.

The $Z_2$ parity and $SU(5)\times U(1)_V \times U(1)_{V^\prime}$
invariant superpotential for the SM matters is 
\begin{eqnarray}
 W = f_u T_{\bf 10} F_{\bf 5^*} \overline{\chi}_{\bf 5^*}
    +f_d T_{\bf 10} T_{\bf 10} H_{\bf 5}
    +f_l F_{\bf 5^*} X_{\bf 1} H_{\bf 5},
\end{eqnarray}
where we omit the indices of generations for simplicity, and $f$'s are
the Yukawa couplings.
Thus, in this model, Dirac neutrino masses and up-type quark masses are
the same each other; $m_u = m_{\nu_D}$, which is in contrast with the
Georgi-Glashow model. 
On the other hand, although down-type quarks and charged leptons acquire
Dirac masses from the same Higgs VEV, there is no relation among their Yukawa
couplings. In other words, although there is no bottom-tau
unification, there are also no wrong relations of $m_s = m_\mu$
and $m_d = m_e$.
Right-handed Majorana masses are given by the following dimension five
operator,
\begin{eqnarray}
 W = \frac{f_R}{M_{pl}} T_{\bf 10} T_{\bf 10}
     \overline{\Sigma}_{\bf 10^*}\overline{\Sigma}_{\bf 10^*}.
\end{eqnarray}
After GUT symmetry breaking, right-handed Majorana neutrinos acquire
the following heavy masses,
\begin{eqnarray}
 M_R \sim \frac{f_R\langle \overline{\Sigma} \rangle^2}{M_{pl}},
\end{eqnarray}
which leads to small neutrino masses through 
the conventional seesaw mechanism.

The superpotential for exotic matters is 
\begin{eqnarray}
 W = y_D\tilde{T}_{\bf 5^*} \tilde{F}_{\bf 5} \phi_{\bf 1}
    +\frac{y_S}{M_{pl}} \tilde{T}_{\bf 1} \tilde{T}_{\bf 1} 
     \overline{\phi}_{\bf 1} \overline{\phi}_{\bf 1}.
\end{eqnarray}
After Higgs fields $\phi$ acquire VEVs, the exotic matters
 become super-heavy and decouple.
Here we assume that VEVs of $\phi$ is slightly smaller than $M$.

Next we discuss the gauge coupling unification.
First of all, notice 
that since the hypercharge assignment of our model is different
from that of Georgi-Glashow model or flipped model, 
the gauge coupling flow is different from those of them.
We normalize the charges and couplings in the following manner (where
$Z$, $V$ and $V^\prime$ are defined in the previous section),
$\tilde{Z} \equiv \sqrt{3/5}Z$. The corresponding coupling is 
$g_{\tilde{Z}} = \sqrt{5/3}g_Z$. This coupling satisfies $g_{\tilde{Z}} = g_5$ 
at the $SU(5)$ unification scale $M_5$. 
$\tilde{V} \equiv 1/2\sqrt{10} V$ and 
$g_{\tilde{V}} = 2\sqrt{10} g_V$. Note here that $tr_r \tilde{V}^2 =
tr_r \tilde {Z}^2$ where $r$ is any $SO(10)$ representation.
Therefore, we expect 
$g_{\tilde{V}} = g_5 = g_{10}$ at the $SO(10)$ unification scale $M_{10}$.
$\tilde{V^\prime} \equiv 1/2\sqrt{6}V^\prime$ and 
$g_{\tilde{V^\prime}} = 2\sqrt{6}g_{V^\prime}$. 
Note here that $tr_r \tilde{V^\prime}^2 = tr_r \tilde{V}^2 = 
tr_r \tilde{Z}^2$ where $r$ is any $E_6$ representation.
We also expect $g_{\tilde{V^\prime}} = g_{10} = g_6$ at the $E_6$
unification scale $M_6$.
We can find the following relation at the $SU(5)$ unification scale,
\begin{eqnarray}
 \frac{3}{5}\frac{1}{\alpha_Y} = 
 \frac{1}{25}\frac{1}{\alpha_5} + \frac{3}{50}\frac{1}{\alpha_{\tilde{V}}}
 + \frac{9}{10}\frac{1}{\alpha_{\tilde{V^\prime}}}, \label{eq:coupling}
\end{eqnarray}
where $\alpha_i$ is defined by $\alpha_i \equiv g_i^2/4\pi$ for any $i$.
Therefore, if $24/25\alpha_5^{-1} \sim 3/50 \alpha_{\tilde{V}}^{-1} 
+ 9/10 \alpha_{\tilde{V^\prime}}^{-1}$ is satisfied at the $SU(5)$
unification scale, the quasi-unification of strong, weak and hypercharge
gauge couplings will realize. 
Since it seems natural that the unification scale of
$SO(10)$ and $E_6$ are not much higher than that of $SU(5)$, this
situation should be possible.
We comment on the ratio $K \equiv \alpha_1/\alpha_5$ at the scale
$M_5$, where $\alpha_1$ is defined by $\alpha_1 \equiv 5\alpha_Y/3$.
The natural condition that $M_5 < M_{10} < M_6$ implies 
$\alpha_{\tilde V}(M_5) < \alpha_5(M_5)$ and 
$\alpha_{\tilde{V^\prime}}(M_{10}) < \alpha_5(M_{10})$.
From these relations and the infrared-free behavior of $U(1)_{V^\prime}$,
we find the following inequality 
$\alpha_{\tilde V}, \alpha_{\tilde V^\prime} < \alpha_5$ at the scale $M_5$.
From both this relation and Eq.~(\ref{eq:coupling}),
we conclude that $K < 1$. 
This is certainly a required relation that MSSM predicts
from the precise experimental values at $M_Z$ scale.

Although the detailed quantitative discussion
and other phenomenological aspects of this model
are interesting and should be pursued, 
that is beyond the scope of this paper.

Final comment is on the Higgs superpotential.
We need, of course, a Higgs superpotential to define the complete model.
We can easily write down 
the most general superpotential which is allowed by the gauge symmetry. 
However, there are some unwanted operators in the most general
superpotential in view of $F$-flatness conditions and/or doublet-triplet
splitting problem.
In order to avoid unwanted operators, it is necessary to impose
some additional symmetry beyond $SU(5) \times U(1)_V \times U(1)_{V^\prime}$.
Although there are some possibilities as an additional symmetry, 
we do not discuss here since they are model dependent.
These possibilities will be investigated elsewhere.

\section{\label{sec:level1}Conclusion}
In this paper we investigated assignments of the 
hypercharge and baryon minus lepton number for particles
in the $E_6$ grand unification model. 
First we reviewed the $SO(10)$ GUT and pointed out that 
as far as $SO(10) \supset U(1)_{B-L}$ is imposed, only the ${\bf 16}$
representation, not ${\bf 10}$, can be a candidate for the SM matters.
Next we studied the $E_6$ GUT and 
it was shown that there are three assignments of hypercharge
and three assignments of $B-L$
which are consistent with the SM. 
Their explicit expressions and detailed properties were given.
We showed that three assignments of hypercharge or $B-L$
are related with each other by the $SU(2)$ subgroup of $SU(3)_R$, 
which is the subgroup of $E_6$.
We also pointed out that any charge assignments are orthogonal to the $SU(2)$
subgroup of $SU(3)_R$.
In particular, we emphasized that the $U(1)_{B-L}$ symmetry in $E_6$
can not be orthogonal to the $SU(3)_R$ symmetry. 
This fact strongly restricts $E_6$ grand unified models
with the gauged $U(1)_{B-L}$ symmetry.
In those of $9$ pairs of charge assignments, we showed that 
$6$ pairs are consistent with the SM. 
These observation indicate that $E_6$ has much potential for constructing
alternative GUT models.

We also proposed a $E_6$-inspired $SU(5)$ GUT model,
``E-twisting flipped model''.
The charge assignments of hypercharge and $B-L$
in this model are different from those of Georgi-Glashow model or usual
flipped model. Since the hypercharge is not a subgroup of
$SU(5)$, the quasi-unification of strong, weak and hypercharge gauge coupling
is predicted. There are also no mass relations between down-type
quarks and charged leptons.

Finally we would like to emphasize that 
it is worth investigating the alternative GUT models 
based on $E_6$ 
to overcome the difficulties of the minimal SUSY $SU(5)$ model.
              
\acknowledgments
We would like to thank J.~Kodaira and T.~Onogi for useful discussion and
their kind comments on our English.


\begin{thebibliography}{99}
\bibitem{SK99}
Y.~Hayato {\it et al}. [Super-Kamiokande Collaboration],
\prl{83}{1999}{1529}.
\bibitem{MP01}
H.~Murayama and A.~Pierce,
\prd{65}{2002}{055009}.
\bibitem{BPS02}
B.~Bajc,~P.~F.~Perez and G.~Senjanovic,
\prd{66}{2002}{075005}.
\bibitem{EW03}
D.~Emmanuel-Costa and S.~Wiesenfeldt,
\hepph{0302272}.
\bibitem{nu-exp}
Y.~Fukuda {\it et al}. [Super-Kamiokande Collaboration],
\prl{81}{1998}{1562};
M.~Apollonio {\it et al}. [CHOOZ Collaboration],
\plb{466}{1999}{415};
S.~Fukuda {\it et al}. [Super-Kamiokande Collaboration],
\prl{86}{2001}{5656};
Q.~R.~Ahmad {\it et al}. [SNO Collaboration],
\prl{89}{2002}{011301}; \prl{89}{2002}{011302};
K.~Eguchi {\it et al}. [KamLAND Collaboration],
\prl{90}{2003}{021802}.
\bibitem{seesaw}
T.~Yanagida, in Proceedings of the 
``{\it Workshop on the Unified Theory and the Baryon Number in the
       Universe}'', Tsukuba, Japan, 1979, 
edited by O. Sawada and A. Sugamoto, KEK Report No. KEK-79-18, p.95; 
M.~Gell-Mann,~P.~Ramond and R.~Slansky, in 
``{\it Supergravity}'', edited by D. Z. Freedman and
	P. van. Nieuwenhuizen (North-Holland, Amsterdam, 1979);
R.~N.~Mohapatra and G.~Senjanovic,
\prl{44}{1980}{912};
S.~L.~Glashow, in
``{\it Quarks and Leptons}'', (Plenum Press, NY, 1980), p.707.
\bibitem{GG74}
H.~Georgi and S.~L.~Glashow,
\prl{32}{1974}{438}.
\bibitem{Barr82}
S.~M.~Barr,
\plb{112}{1982}{219}.
\bibitem{flipped}
J.~P.~Derendinger,~J.~E.~Kim and D.~V.~Nanopoulos,
\plb{139}{1984}{170};
I.~Antoniadis,~J.~Ellis,~J.~S.~Hagelin and D.~V.~Nanopoulos,
\plb{194}{1987}{231};
J.~Ellis,~J.~L.~Lopez and D.~V.~Nanopoulos,
\plb{371}{1996}{65}.
\bibitem{Reflipped}
J.~Ellis,~D.~V.~Nanopoulos and J.~Walker,
\plb{550}{2002}{99};
D.~V.~Nanopoulos,
\hepph{0211128}.
\bibitem{E6}
F.~G\"ursey,~P.~Ramond and P.~Sikivie,
\plb{60}{1976}{177};
Y.~Achiman and B.~Stech, 
\plb{77}{1978}{389}
\bibitem{BK99}
M.~Bando and T.~Kugo, 
\ptp{101}{1999}{1313}.
\bibitem{E6model}
M.~Bando,~T.~Kugo and K.~Yoshioka,
\ptp{104}{2000}{211};
M.~Bando and N.~Maekawa,
\ptp{106}{2001}{1255}.
\bibitem{Hyper}
See also,
D.~London and J.~L.~Rosner, 
\prd{34}{1986}{1530}.
\bibitem{Ma87}
E. Ma,
\prd{36}{1987}{274}.
\bibitem{Ma96}
E. Ma,
\plb{380}{1996}{286}.
\end{thebibliography}
\end{document}